\documentclass[12pt]{iopart} 
\usepackage{graphicx}
\usepackage{amssymb}
\usepackage{ulem}  
\usepackage{color} 
\usepackage{epstopdf}
\usepackage{cite}

\begin{document}

\title{Effect of packing fraction on the jamming of granular flow through small apertures}

\author{Rodolfo Omar U\~nac$^1$, Ana Mar\'{\i}a Vidales$^1$ and Luis A. Pugnaloni$^2$}
\address{$^1$ Departamento de F\'{\i}sica, Instituto de F\'{\i}sica Aplicada (UNSL-CONICET), Universidad Nacional de San Luis, Ej\'ercito de los Andes 950, D5700HHW San Luis, Argentina.\\
$^2$ Instituto de F\'{\i}sica de L\'{\i}quidos y Sistemas Biol\'{o}gicos (CONICET La Plata, UNLP), Calle 59 Nro 789, 1900 La Plata, Argentina.}
\ead{runiac@unsl.edu.ar (R O U\~nac), avidales@unsl.edu.ar (A M Vidales), luis@iflysib.unlp.edu.ar (L A Pugnaloni)}

\begin{abstract}
We investigate the flow and jamming through small apertures of a column of granular disks via a pseudo-dynamic model. We focus on the effect that the preparation of the granular assembly has on the size of the avalanches obtained. Ensembles of packings with different mean packing fractions are created by tapping the system at different intensities. Surprisingly, packing fraction is not a good indicator of the ability of the deposit to jam a given orifice. Different mean avalanche sizes are obtained for deposits with the same mean packing fraction that were prepared with very different tap intensities. It has been speculated that the number and size of arches in the bulk of the granular column should be correlated with the ability of the system to jam a small opening. We show that this correlation, if exists, is rather poor. A comparison between bulk arches and jamming arches (i.e., arches that block the opening) reveals that the aperture imposes a lower cut-off on the horizontal span of the arches which is greater than the actual size of the opening. This is related to the fact that blocking arches have to have the appropriate orientation to fit the gap between two piles of grains resting at each side of the aperture.
\end{abstract}

\maketitle

\section{Introduction}

The flow of granular materials through apertures is commonplace in a variety of industrial applications. Studies in this respect can be separated into two major areas: (1) continuous flow, and (2) jamming. Continuous granular flow is observed for dry non-cohesive materials if the size of the aperture is large (typically above five grain diameters for spherical particles). Jamming is observed whenever the opening is smaller, which requires the input of external perturbations in order to restart the flow. The cause of jamming is the formation of a blocking arch. We distinguish two type of arches: blocking arches (or jamming arches) and bulk arches. Blocking arches form at the orifice during discharge and prevent the flow. Bulk arches form at any place inside the packing during the dynamical process that leads the grains to reach mechanical equilibrium. Arches are set of particles that are mutually stable. The removal of any particle in the arch leads to the destabilization of the others.

A number of studies have considered the jamming of an aperture during the discharge of grains. These include experimental studies on two-dimensional (2D) hoppers using circular grains \cite{to1,to2} and three-dimensional (3D) silos using spherical and non-spherical particles \cite{zuriguel1,zuriguel2}, numerical simulations using discrete element methods in 2D \cite{magalhaes,perez}, experiments with quasi-2D silos and spherical grains \cite{janda}, 3D vibrated silos \cite{mancok} and experiments with tilted \cite{sheldon} and wedge-shaped hoppers \cite{saraf}. Also, the properties of blocking arches and bulk arches have been considered in the past \cite{to1,garcimartin,pugnaloni2,pugnaloni3,arevalo,mehta2}. However, none of these studies have considered the effect of the preparation of the granular column prior to the discharge. It is known that the number and size of arches inside a granular assembly are dependent on the packing fraction. Therefore, it is expected that the jamming of columns prepared at different packing fractions may occur with different probability. A related issue is the question of to what extent the arches formed in the bulk of the system are comparable with the arches that effectively block the aperture during drainage. 

In this paper we use a 2D pseudo-dynamic simulation scheme previously developed by Manna and Khakhar \cite{manna1,manna2} to study the effect of the initial packing fraction on the jamming probability and the correlation between bulk arches and blocking arches. The jamming probability is directly connected with the mean size of the avalanches \cite{janda}. An avalanche is the flow of grains that occurs between the initiation of the discharge and the arrest of the flow due to the formation of a blocking arch \cite{zuriguel1}. We will show that there is a strong dependence of the size of the avalanches with the packing fraction. However, there is not a monotonic relation between these two quantities. We also find that there is a poor correlation between the size of arches in the bulk and the size of the avalanches. A comparison between bulk arches and jamming arches reveals that the aperture not only imposes a cut-off on arches of horizontal span below the opening size, but also prevents the formation of some blocking arches that, in principle, are wide enough to induce jamming.

\section{The pseudo-dynamic algorithm}

Our simulations are based on an algorithm for inelastic massless hard disks designed by Manna and Khakhar \cite{manna1,manna2}. This is a pseudo-dynamics that consists in small falls and rolls of the grains until they come to rest by contacting other particles or the system boundaries. We use a container formed by a flat base and two flat vertical walls. No periodic boundary conditions are applied. 

The deposition algorithm consists in choosing a disk in the system and allowing a free fall of length $\delta$ if the disk has no supporting contacts, or a roll of arc-length $\delta$ over its supporting disk if the disk has one single supporting contact \cite{manna1,manna2,pugnaloni6}. Disks with two supporting contacts are considered stable and left in their positions. If in the course of a fall of length $\delta$ a disk collides with another disk (or the base), the falling disk is put just in contact and this contact is defined as its \textit{first supporting contact}. Analogously, if in the course of a roll of length $\delta$ a disk collides with another disk (or a wall), the rolling disk is put just in contact. If the \textit{first supporting contact} and the second contact are such that the disk is in a stable position, the second contact is defined as the \textit{second supporting contact}; otherwise, the lowest of the two contacting particle is taken as the \textit{first supporting contact} of the rolling disk and the \textit{second supporting contact} is left undefined. If, during a roll, a particle reaches a lower position than the supporting particle over which it is rolling, its \textit{first supporting contact} is left undefined (in this way the particle will fall vertically in the next step instead of rolling underneath the first contact). A moving disk can change the stability state of other disks supported by it, therefore, this information is updated after each move. The deposition is over once each particle in the system has both supporting contacts defined or is in contact with the base (particles at the base are supported by a single contact). Then, the coordinates of the centers of the disks and the corresponding labels of the two supporting particles, wall, or base, are saved for analysis.

\begin{figure}[t]
\begin{center}
\includegraphics[width=0.5\columnwidth]{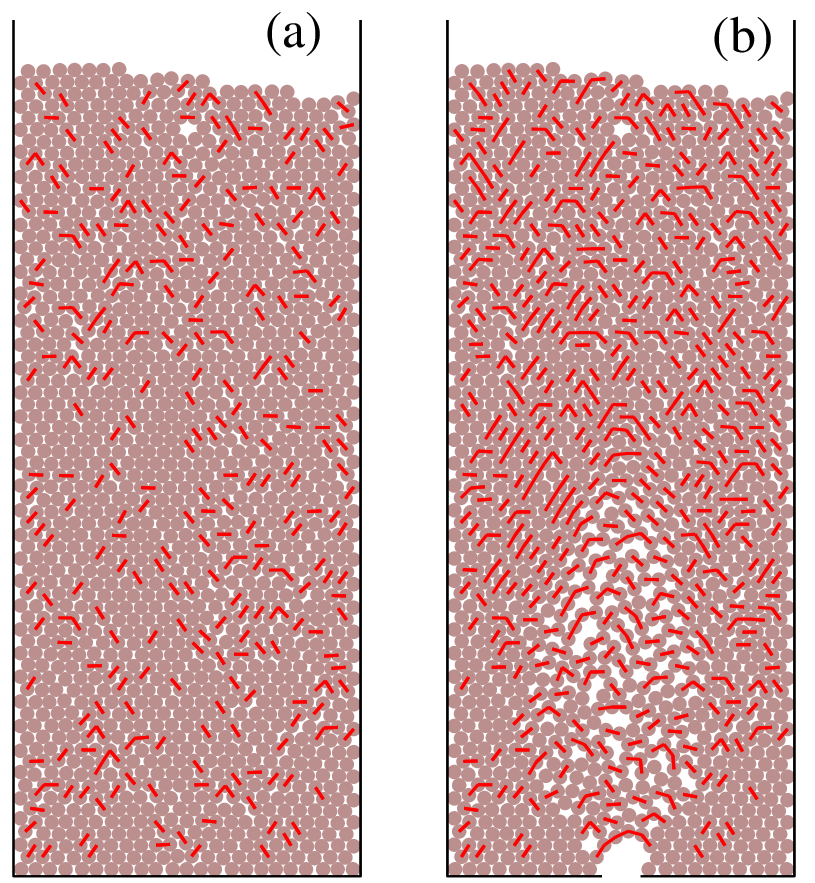}
\caption {(Color online). Sample configurations of the granular column for the steady state corresponding to $\Gamma=0.39$ before (a) and after (b) discharging through and opening of width $D=2.75$. The red segments indicate the arches in the system.}
\label{fig1}
\end{center}
\end{figure}

An important point in these simulations is the effect that the parameter $\delta$ has in the results since particles do not move simultaneously but
one at a time. One might expect that in the limit $\delta \rightarrow 0$ we should recover a fairly "realistic" dynamics for fully inelastic non-slipping disk dragged downwards at constant velocity. This should represent particles deposited in a viscous medium or carried by a conveyor belt. Although this dynamics contrasts with the dynamics of dry granulates, experiments on the jamming of fully submerged grains in gels \cite{roussel} have shown remarkable similarities with the more widely available data on dry systems. We chose $\delta=0.0062 d$ (with $d$ the particle diameter) since we have observed that for smaller values of $\delta$ results are indistinguishable from those obtained here \cite{pugnaloni6}.

The pseudo-dynamics approach has been chosen in view of the low CPU time demanded by this scheme. In this study, we need to prepare a large number of  samples (through tapping) with a given packing fraction and then trigger discharges for different aperture sizes. More realistic simulations such as granular dynamics (or discrete element method) would require a much higher computational effort. In spite of the simplifications of the pseudo-dynamics, it has been shown that results on tapping agree qualitatively with granular dynamics simulations \cite{pugnaloni4}. 

\section{Initial packings}

In order to study the effect of packing fraction on the jamming of the flow through an aperture, we first need to prepare packings at reproducible packing fractions. To achieve this, several techniques can be applied. For example, the sequential deposition of grains submerged in a viscous liquid yield reproducible packing fractions that can be tuned by changing the friction coefficient of the particles or the density mismatch \cite{farrell}. We have chosen another well known technique to generate reproducible ensembles of packings: tapping. Nowak et al. have shown that an appropriate tapping protocol can lead to reproducible states in the sense that an ensemble of configurations with well defined mean packing fraction is recovered if the same protocol is followed irrespective of the initial state \cite{nowak}. This has been more carefully discussed by Ribi\`{e}re et al. \cite{ribiere}. Dijksman et al. showed how different states can be obtained not only by changing the tap intensity but also by changing the tap duration \cite{dijksman}. A similar effect was investigated by Pica Ciamarra et al. in submerged samples where a fluid pulse is used to excite the granular column \cite{ciamarra}. Hence, we use a simulated tapping protocol (see below) to generate sets of initial configurations that have well defined mean packing fractions.

\begin{figure}[t]
\begin{center}
\includegraphics[width=0.5\columnwidth]{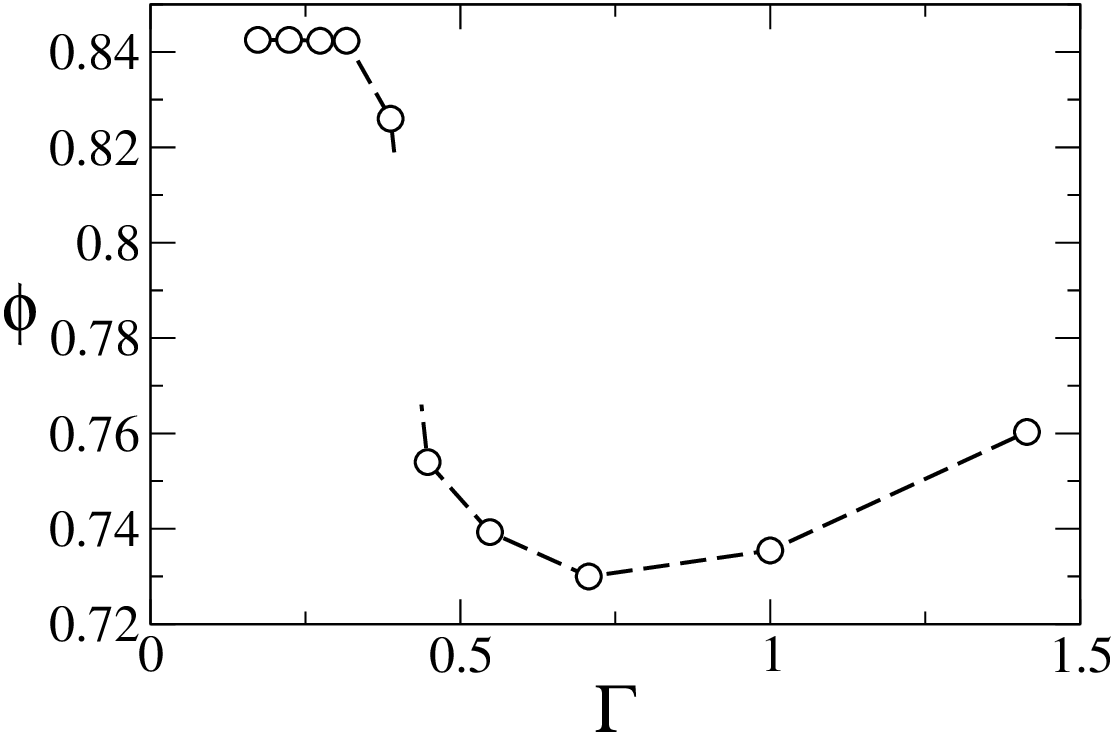}
\caption {Mean packing fraction $\phi$ in the steady state of the tapping protocol as a function of the tap intensity $\Gamma$.}
\label{fig2}
\end{center}
\end{figure}

The simulations are carried out in a rectangular box of width $24.78 d$ containing $1500$ equal-sized disks of diameter $d$. Initially, disks are placed at random in the simulation box (with no overlaps) and deposited using the pseudo-dynamic algorithm. Once all the grains come to rest, the system is expanded in the vertical direction and randomly shaken to simulate a vertical tap. Then, a new deposition cycle begins. After many taps of given amplitude, the system achieves a steady state where all characterizing parameters fluctuate around equilibrium values independently of the previous history of the granular bed. The existence of such ``equilibrium'' states has been previously reported in experiments \cite{ribiere}. 

The tapping of the system is simulated by multiplying the vertical coordinate of each particle by a factor $A$ (with $A>1$). Then, the particles are subjected to several (about $20$) Monte Carlo loops where positions are changed by displacing particles a random length $\Delta r$ uniformly distributed in the range $0<\Delta r<A-1$. New configurations that correspond to overlaps are rejected. This disordering phase is crucial to avoid particles falling back again into the same positions. Moreover, the upper limit for $\Delta r$ (i.e. $A-1$) is deliberately chosen so that a larger tap promotes larger random changes in the particle positions. The expansion amplitude $A$ ranges from $1.03$ up to $3.0$. Following Refs. \cite{philippe,pugnaloni4} we quantify the tap intensity by the parameter $\Gamma=\sqrt{A-1}$. For each value of $\Gamma$ studied, $10^{3}$ taps are carried out for equilibration followed by $5 \times 10^3$ taps for production. $500$ deposited configurations are stored which are obtained by saving every $10$ taps during the production run after equilibration. These deposits will be used later as initial conditions for the discharge and flow through an opening.

The deposited configurations are analyzed in search of bulk arches. We first identify all mutually stable particles ---which we define as directly connected--- and then we find the arches as chains of connected particles. Two disks A and B are mutually stable if A is the left supporting particle of B and B is the right supporting particle of A, or viceversa. We measure the total number of arches, arch size distribution $n(k)$, and the horizontal span distribution of the arches $n_{k}(x)$. The latter is the probability density of finding an arch consisting of $k$ disks with horizontal span between $x$ and $x+dx$. The horizontal span (or lateral extension) is defined as the projection onto the horizontal axis of the segment that joins the centers of the right-end disk and the left-end disk in the arch. In Fig.~\ref{fig1}(a), we show an example of a deposited configuration with arches indicated by segments (for a description of Fig.~\ref{fig1}(b) see next section). Notice that the pseudo-dynamics mimics the behavior of disks that roll without slipping. This corresponds to a system with infinite static friction which is expected to yield a large number and variety of arches. The arch structure of frictionless systems may differ significantly from the one seen in Fig. \ref{fig1}(a). However, simulations with realistic discrete element methods with finite friction yield similar structures \cite{arevalo}.

In Fig.~\ref{fig2}, we present the steady state mean packing fraction, $\phi$, of our granular deposits as a function of $\Gamma$. There exists a rather sharp decrease of $\phi$ as the tapping intensity is increased followed by a minimum and a very smooth increase. The sharp drop of $\phi$ is associated to a discontinuous order-disorder transition previously reported for this model \cite{pugnaloni6} and also observed in granular dynamics of polygonal grains \cite{carlevaro}. The appearance of the minimum packing fraction as a function of tap intensity has been reported for several models (including a frustrated lattice gas model \cite{gago}, a Monte Carlo type deposition \cite{pugnaloni4} and a realistic discrete element method simulation \cite{pugnaloni4}) and in experiments of tapping with a quasi-2D system \cite{pugnaloni1,pugnaloni5}. For the model we use in this paper, the minimum $\phi$ has been shown to exist even if bidisperse systems are considered \cite{vidales}.

\begin{figure}[t]
\begin{center}
 \includegraphics[width=0.5\columnwidth]{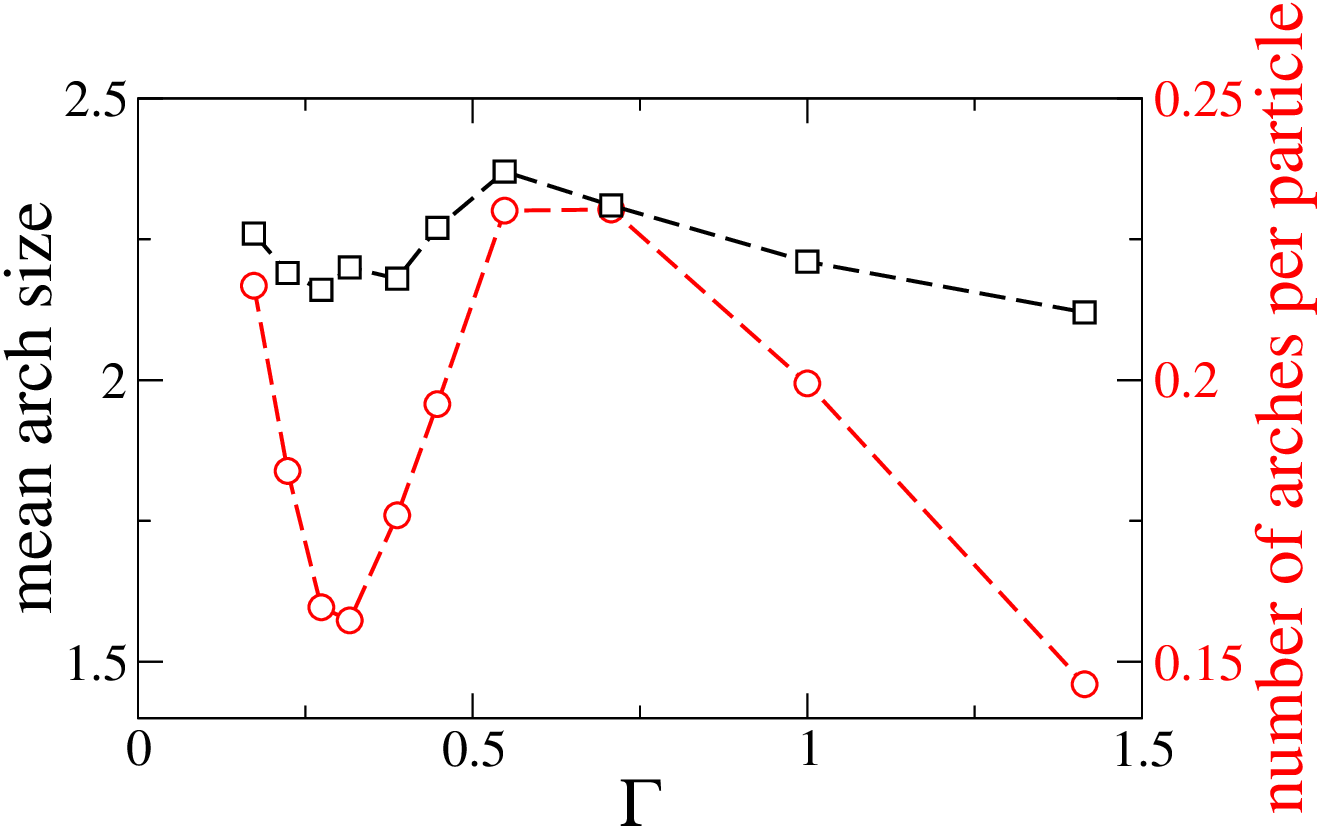}
\caption {(Color online). Number of arches per particle (red circles) and mean size of the arches in terms of the number of grains involved in an arch (black squares) as a function of $\Gamma$.}
\label{fig3}
\end{center}
\end{figure}

The minimum in $\phi$ is caused to a large extent by the formation of arches \cite{pugnaloni4}. Figure \ref{fig3} shows the number and mean size of arches found in the system as a function of $\Gamma$. When $\Gamma$ is increased considerably (above $0.7$), every tap expands the assembly in such a way that particles get well apart from each other. During deposition, particles will reach the free surface of the bed almost sequentially (one at a time), reducing the chances of mutual stabilization. Therefore, arches are less probable to form as $\Gamma$ increases and so $\phi$ must grow since fewer voids get trapped. Indeed, we see in Fig.~\ref{fig3} that the number and size of arches decrease at large $\Gamma$ ($\gtrsim0.7$) for increasing tap intensities. Eventually, for very large $\Gamma$, no arches are formed after each tap and $\phi$ will reach a limiting value. At lower $\Gamma$ ($0.4\lesssim\Gamma\lesssim0.7$), the free volume injection due to a tap creates very narrow gaps between particles. For a given arch to grow by the insertion of a new particle, it is necessary to create a gap between two particles in the existing arch where the new particle can fit in. This explains why increasing $\Gamma$ will promote the formation of larger arches and reduce $\phi$ in this regime. For very low tapping intensities ($\Gamma\lesssim0.3$), we find a rather constant $\phi$. However, the number and size of the arches decrease with $\Gamma$ (see Fig.~\ref{fig3}). This would imply that a maximum in the packing fraction should be observed at such light tapping. This maximum has been recently reported in other models \cite{carlevaro,rosato}, but is not present in our pseudo-dynamics. Notice that the number and size of the arches give only a rough indication for the free volume in the sample since the actual shape of the arches will also be important.

\section{Flow and jamming}

For each deposit generated as described in the previous section, we trigger a discharge by opening an aperture of width $D$ relative to the diameter $d$ of the disk in the center of the base of the containing box. Grains will flow out of the box following the pseudo-dynamics until a blocking arch forms or until the entire system is discharged (with the exception of two piles resting on each side of the aperture). During the dynamics, disks that reach the bottom and whose centers lie on the interval that defines the opening will fall vertically (even if the surface of the disk touches the edge of the aperture). This prevents the formation of arches with end disks sustained by the vertical edge of the orifice. Although such arrangements happen in real experiments, they are uncommon \cite{garcimartin}. After each discharge, we record the size of the avalanche (i.e. the number of grains flowed out) and the final arrangement of the grains left in the box. Averages are taken over 500 discharges for each value of $\Gamma$ used to prepare the initial packings.

One single discharge attempt is carried out for each initial deposit. This allows us to assure that the initial preparation of the pack belongs to the ensemble of deposits corresponding to the steady state of the particular tap intensity chosen. In many experiments and most industrial applications, discharges are triggered one after another from the same deposit without preparing the system in the initial condition again \cite{zuriguel1, zuriguel2, janda, mancok}. However, some experiments do fill the container anew before each discharge \cite{to1,to2,pournin}. We can see in Fig.~\ref{fig1}(b) an image of the system after a discharge that resulted in a jam. It is clear that the structure of the packing is greatly affected by the partial discharge in our simulations. Therefore, these final structures are not used for new discharges.

\begin{figure}[t]
\begin{center}
\includegraphics[width=0.9\columnwidth]{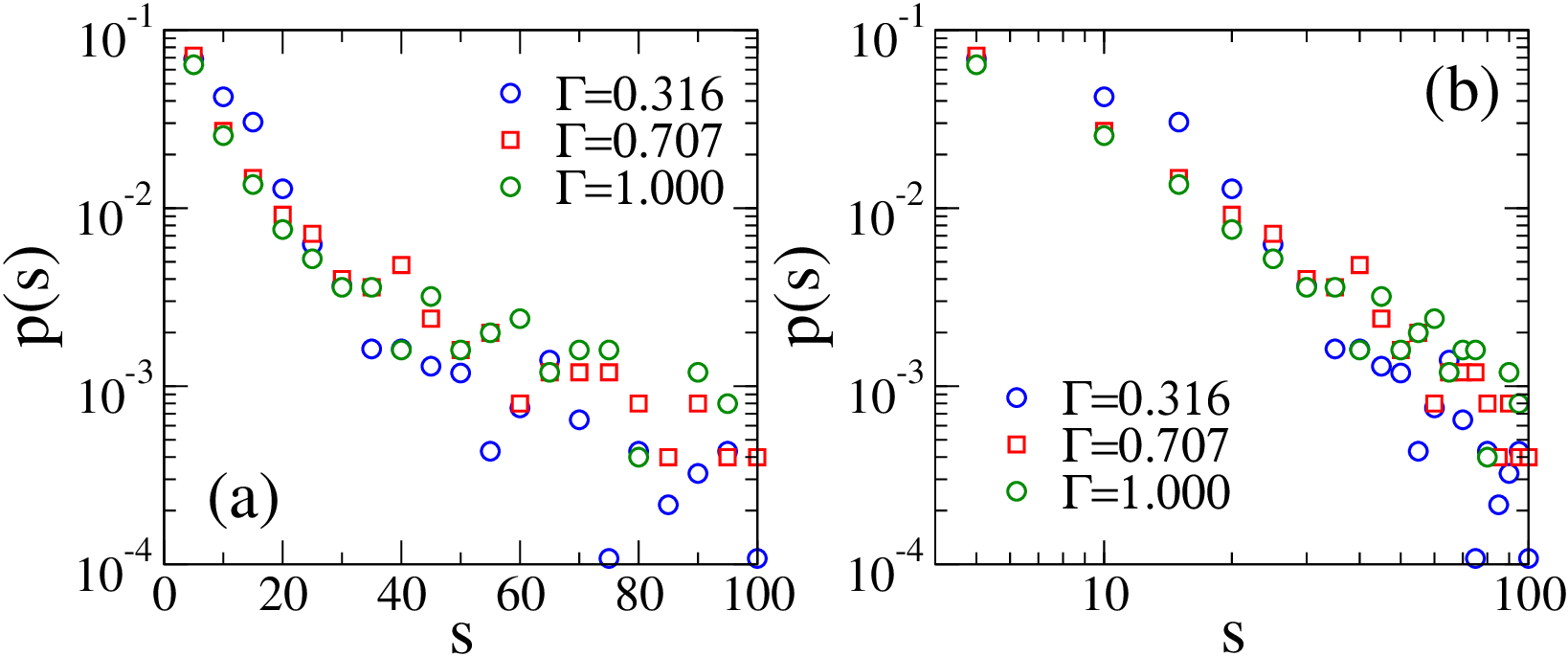}
\caption {(Color online). Avalanche size distribution $p(s)$ for several values of $\Gamma$ for $D=2.25$. (a) Semilog plot, (b) log-log plot.}
\label{fig4}
\end{center}
\end{figure}

In Fig.~\ref{fig4} we plot the avalanche size distribution $p(s)$ for a few values of $\Gamma$ and $D=2.25$. We obtain this by counting the number of grains $s$ that flow in each of the $500$ discharges corresponding to each initial deposit generated for each $\Gamma$. An exponential tail in $p(s)$ has been observed in several previous studies, both two-dimensional \cite{janda,to2,perez,magalhaes} and three-dimensional \cite{zuriguel1,zuriguel2}. Manna and Herrmann \cite{manna3}, using the same model, found avalanches with a power law distribution. Notice however, that in Ref. \cite{manna3} the authors trigger one avalanche after the other by simply removing a grain of the blocking arch. This minute perturbation to trigger avalanches may induce strong correlations between successive discharges in contrast with the strong rearrangements induced in most experiments. Based in our limited number of discharges, we are unable to assert if an exponential or a power-law decay is at play in our simulations (see log-lin and log-log plots in Fig. \ref{fig4}). Although we report $p(s)$ up to $s=100$, larger avalanches of up to $1000$ disks are observed, whose statistics can be largely affected by the finite size of the system ($N=1500$).

\begin{figure}[t]
\begin{center}
\includegraphics[width=0.5\columnwidth]{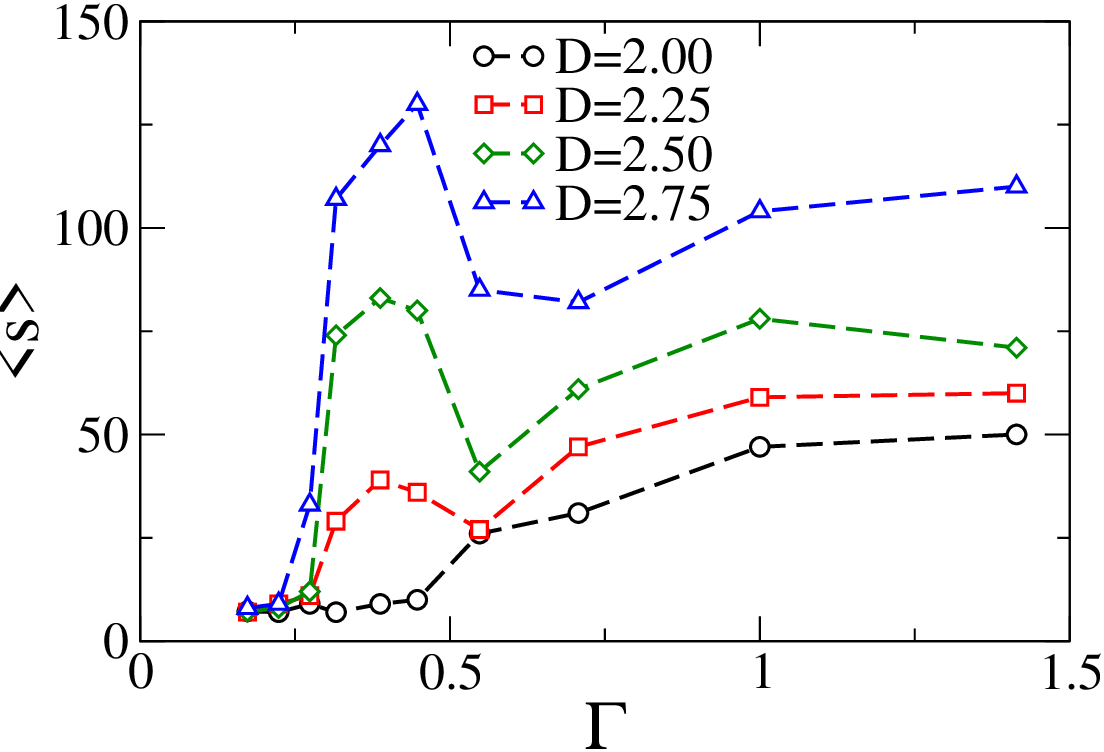}
\caption {(Color online). Mean avalanche size $\langle s \rangle$ as a function of $\Gamma$ for several sizes of the aperture $D$. Notice that $\langle s \rangle$ is affected to a large extent by the rare large avalanches not reported in Fig. \ref{fig4}.}
\label{fig5}
\end{center}
\end{figure}

The mean avalanche size, $\langle s \rangle$, as a function of $\Gamma$ is shown in Fig.~\ref{fig5} for various aperture sizes. As it can be expected, $\langle s \rangle$ increases if $D$ increases. As we can see, for small apertures, $\langle s \rangle$ grows monotonically as $\Gamma$ increases. However, for $D>2.0$, the mean avalanche size presents a local maximum and a local minimum as a function of $\Gamma$.

It has been speculated \cite{mehta1,pugnaloni2} that the size of the avalanches can be connected with the arches inside the granular deposit. Although arches in the initial configuration are not dragged to the aperture during flow since arches actually break and form all the time in the process, it is believed that the ability of the system to form arches in the initial deposit is connected with the ability to form blocking arches during flow. Indeed, the features observed in Fig. \ref{fig5} are somewhat correlated with the number and size of the bulk arches. As $\Gamma$ is increased from the lower values, the size of the arches remains initially rather constant (whereas the number of arches decreases, Fig. \ref{fig3}). This reduction in the number of arches leads to a smaller jamming probability and a rapid increase in the size of the avalanches (see Fig. \ref{fig5}). This regime ends when the sharp drop in $\phi$ ends ($\Gamma \approx 0.4$). For larger $\Gamma$ and up to the packings with minimum packing fractions (i.e. $0.4\lesssim \Gamma \lesssim 0.7$), the number and size of the arches increase. As a consequence, $\langle s \rangle$ decreases due to the increased likelihood of jamming. Finally, for $\Gamma \gtrsim 0.7$, the number and size of the arches fall and a corresponding increase of the avalanche sizes is observed. From these observations we can assert that the number and size of arches in a given packing give an overall indication of the chances that the system will jam if it is left to flow through a small aperture.  

\begin{figure}[t]
\begin{center}
\includegraphics[width=0.9\columnwidth]{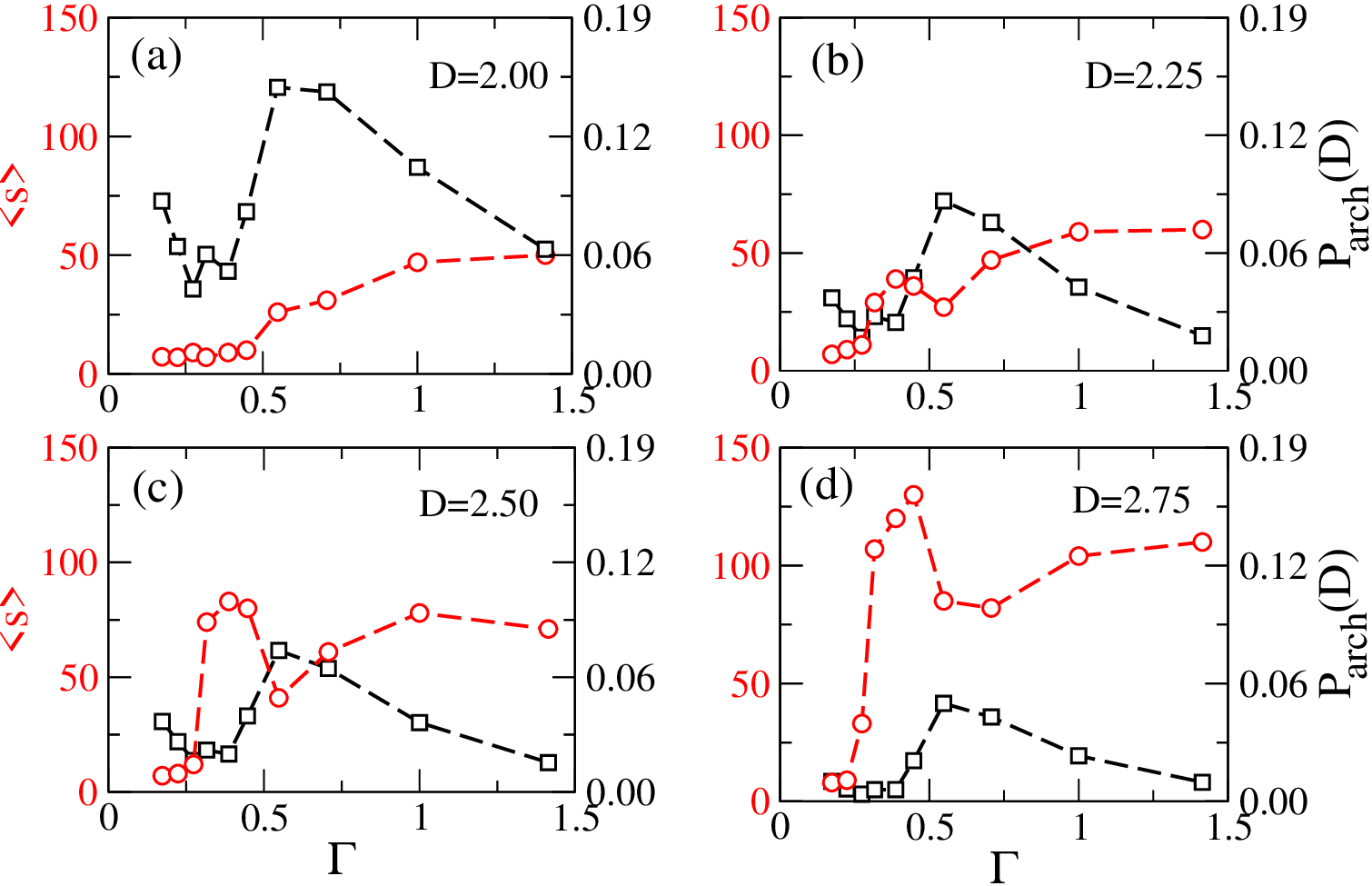}
\caption {(Color online). Mean avalanche size $\langle s \rangle$ as a function of $\Gamma$ and probability $P_\mathrm{arch}(D)$ of finding an arch of horizontal span $D-1$ for several apertures: (a) $D=2.0$, (b) $D=2.25$, (c) $D=2.5$, (d) $D=2.75$.}
\label{fig6}
\end{center}
\end{figure}

Notwithstanding the previous analysis, the size of the avalanches is not only dependent on the preparation protocol ---defined in our case by $\Gamma$--- but also on the actual size of the outlet imposed. For example, as we mentioned, for $D=2.0$ the mean avalanche size does not present the maximum and minimum suggested by the number and size of arches. In order to take into account the effect of the size of the aperture we measure the probability $P_\mathrm{arch}(D)$ of finding an arch in the bulk of the deposits wide enough to block a given aperture $D$. We measure the horizontal span of each arch as the projection on the horizontal axis of the segment that joins the centers of the end particles of the arch. An arch of span $x$ can jam an opening of width $x+d$ (with $d$ the diameter of a grain). In this analysis we include the grains that do not form arches, which can jam any orifice  with $D\leq 1$. In Fig.~\ref{fig6} we plot $P_\mathrm{arch}(D)$ as a function of $\Gamma$ and compare with the corresponding $\langle s \rangle$. Overall, the probability of finding an arch wide enough to block an aperture of size $D$ decreases with $D$ in correspondence with the overall increase of $\langle s \rangle$. However, for a given $D$, the dependence with $\Gamma$ does not show a clear anti-correlation between the probability and the mean avalanche size. This implies that the bulk arches can give only a rough indication of the eventual size of the avalanches that would discharge if an aperture is opened.

It is important to mention that we have always opened the aperture at the center of the bottom of the container. The ordering observed for $\Gamma < 0.5$ (see Fig. \ref{fig1}) suggests that an effect related to the relative position of the aperture and the first layer of grains may be expected. The main effect would be related to the fact that particles just at the edges of the orifice do not flow in the pseudo-dynamics and these will produce a reduced  effective aperture. If all discharges start from an initial packing so ordered that grains at the first layer always sit on the same horizontal positions, the mean avalanche size would depend on the horizontal position of the aperture. When this happens, one observes oscillations in the avalanche size as a function of the aperture size \cite{pournin}. However, this does not happen in our simulations as we can see in Fig. \ref{fig5}. This effect is observed only for highly ordered structures obtained with frictionless particles \cite{pournin}. Our particles model non-slipping grains and the small deviations in position of the disks of the first layer with respect to a truly crystalline structure are sufficiently large to mask any systematic effect due to ordering that may require a detailed study on the position chosen for the aperture.

\section{Effect of packing fraction}

It is generally believed that packing fraction is a good parameter to characterize many properties of a granular bed \cite{bratberg,gravish}.  Results obtained for deposits prepared at a given $\phi$ are not necessarily general and must be repeated for different packing fractions. However, this does not mean that packing fraction is the only or the main factor that can affect the results. In Table~\ref{table1} we present part of the data of Fig. \ref{fig5} but ordered by mean packing fractions $\phi$ corresponding to the steady state obtained for each given $\Gamma$ (see Fig. \ref{fig2}). As we can see, the mean avalanche size depends on $\phi$ in a non trivial way. At the highest $\phi$, obtained by light tapping, the mean avalanche size can range from a few grains to a hundred grains depending on the value of $\Gamma$ used to create the packings. The larger the aperture $D$, the wider the range in $\langle s \rangle$ within this regime where the system is rather ordered but the number and size of the arches fall with increasing $\Gamma$ (see Fig. \ref{fig3}). For low $\phi$, deposits with similar packing fractions but created with low and high tap intensities display different values of $\langle s \rangle$ for any given $D$.   

As we can see, deposits with the same $\phi$ can present different values of $\langle s \rangle$. Therefore, the steady state ensembles of packings with equal $\phi$ obtained by tapping may behave differently. This has been pointed out in Refs. \cite{pugnaloni5,pugnaloni1} where steady states with the same $\phi$ but bearing different stresses were obtained through tapping. In our simulations, forces are not calculated and therefore the stress tensor cannot be obtained. However, a clear difference in the response of the granular columns with same $\phi$ is observed in the sense that avalanches are, in average, of different size.

\begin{table}

\begin{center}
 \begin{tabular}{cc|cccc}
\hline
\hline
  & &  & & $\langle s \rangle$  \\
\hline
$\phi$ & $\Gamma$ & $D=2.00$ & $D=2.25$ &$D=2.50$ & $D=2.75$ \\
\hline
0.8425 & 0.173 & 7.2 & 7 & 7.1 & 8 \\
0.8425 & 0.224 & 7 & 9 & 8 & 9 \\
0.8424 & 0.274 & 9 & 11 & 12 & 33 \\
0.8424 & 0.316 & 7 & 29 & 74 & 107 \\
\hline
0.7393 & 0.548 & 26 & 27 & 41 & 85 \\          
0.7354 & 1.000 & 47 & 59 & 78 & 104 \\                 
\hline
0.7540 & 0.447 & 10 & 36 & 80 & 130 \\
0.7603 & 1.414 & 50 & 60 & 71 & 110 \\         
\hline
\hline
 \end{tabular} 
\end{center}
\caption{Mean avalanche size $\langle s \rangle$ for different tap intensities $\Gamma$ that yield similar packing fractions $\phi$.}
\label{table1}
\end{table} 

\section{Connection between bulk arches and jamming arches}

Although we have shown in the previous section that arches found in the bulk of the granular packing give a rough indication as to whether the system would be more or less likely to jam, arches actually formed at the aperture during discharge are different. Detailed studies of such blocking arches have been reported for two-dimensional experimental setups \cite{garcimartin,to3}. Since we have access to both bulk and blocking arches in our simulations, we compare a few properties and discuss on the implications for the jamming probability.

\begin{figure}[t]
\begin{center}
\includegraphics[width=0.5\columnwidth]{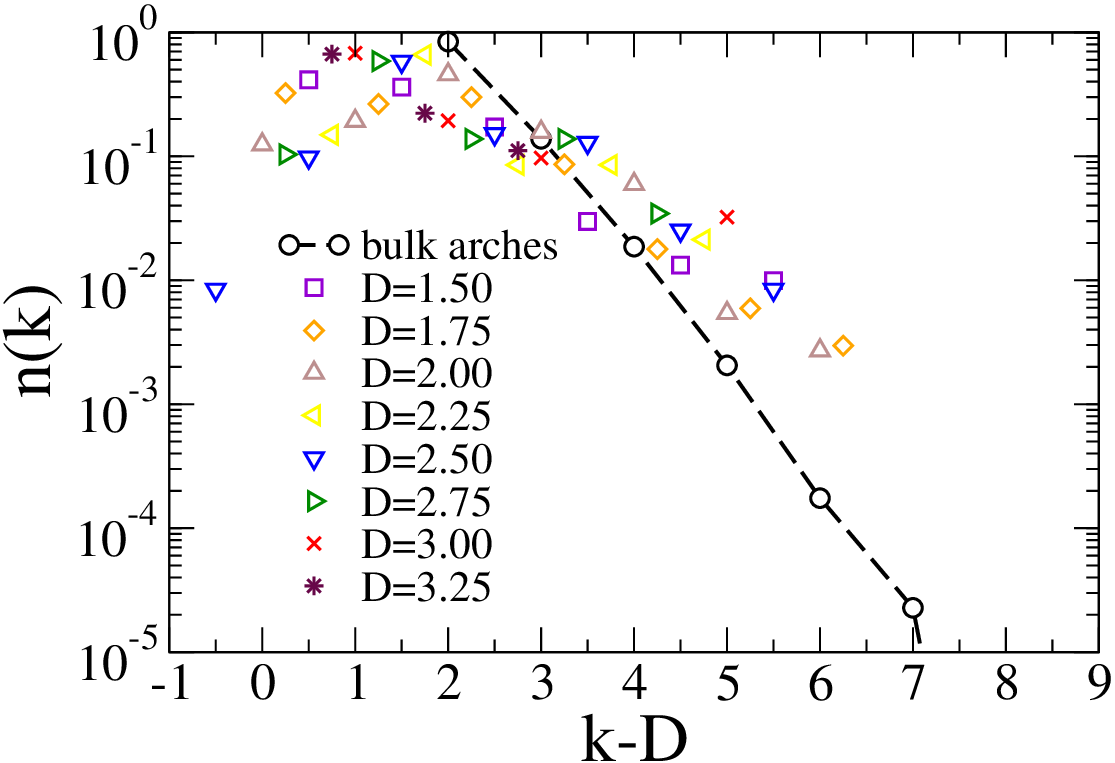}
\caption {(Color online). Distribution $n(k)$ of arch sizes for $\Gamma=0.387$.  The black symbols correspond to arches found in the bulk prior to the discharge, the color data correspond to the blocking arches for different values of $D$ as indicated in the legend. Notice that the horizontal axis is shifted by the size of the aperture $D$ for each set of data.}
\label{fig7}
\end{center}
\end{figure}

In Fig.~\ref{fig7} we compare the arch size distribution $n(k)$ for the bulk arches and for the jamming arches for different values of $D$. $n(k)$ is the probability of finding an arch of $k$ grains ($k\geq 2$). For bulk arches, $n(k)$ is calculated as the number of arches of $k$ disks found in all initial packings divided by the total number of arches (i.e., summing for all $k\geq 2$). For jamming arches, $n(k)$ is calculated as the number of discharges that led to a blocking arch of $k$ disks divided by the total number of discharges (discharges that ended without producing a jam were not considered). The plot is presented as a function of $k-D$ since this produces a collapse of the curves by subtracting a quantity ($D$) proportional to the lower cut-off imposed by the orifice on the arch sizes (see Ref. \cite{garcimartin}). The fact that all the normalized histograms fall on the same curve indicates that the nature of the arches that jam the aperture is the same for small and big orifices. In general, an exponential tail is observed and a cut off for small arches is imposed by the orifice. The large arches generally form upstream supported by two stationary piles resting on both sides of the opening. It is clear from Fig. \ref{fig7} that blocking arches tend to be larger than bulk arches (even after the correction due to the cut-off imposed by $D$). We believe this is due to the fact that many small arches that are stable in the bulk thanks to the many neighbors cannot accommodate in the conical shaped funnel created by the two stationary piles. This is best demonstrated in the next paragraph.

\begin{figure}[t]
\begin{center}
\includegraphics[width=0.9\columnwidth]{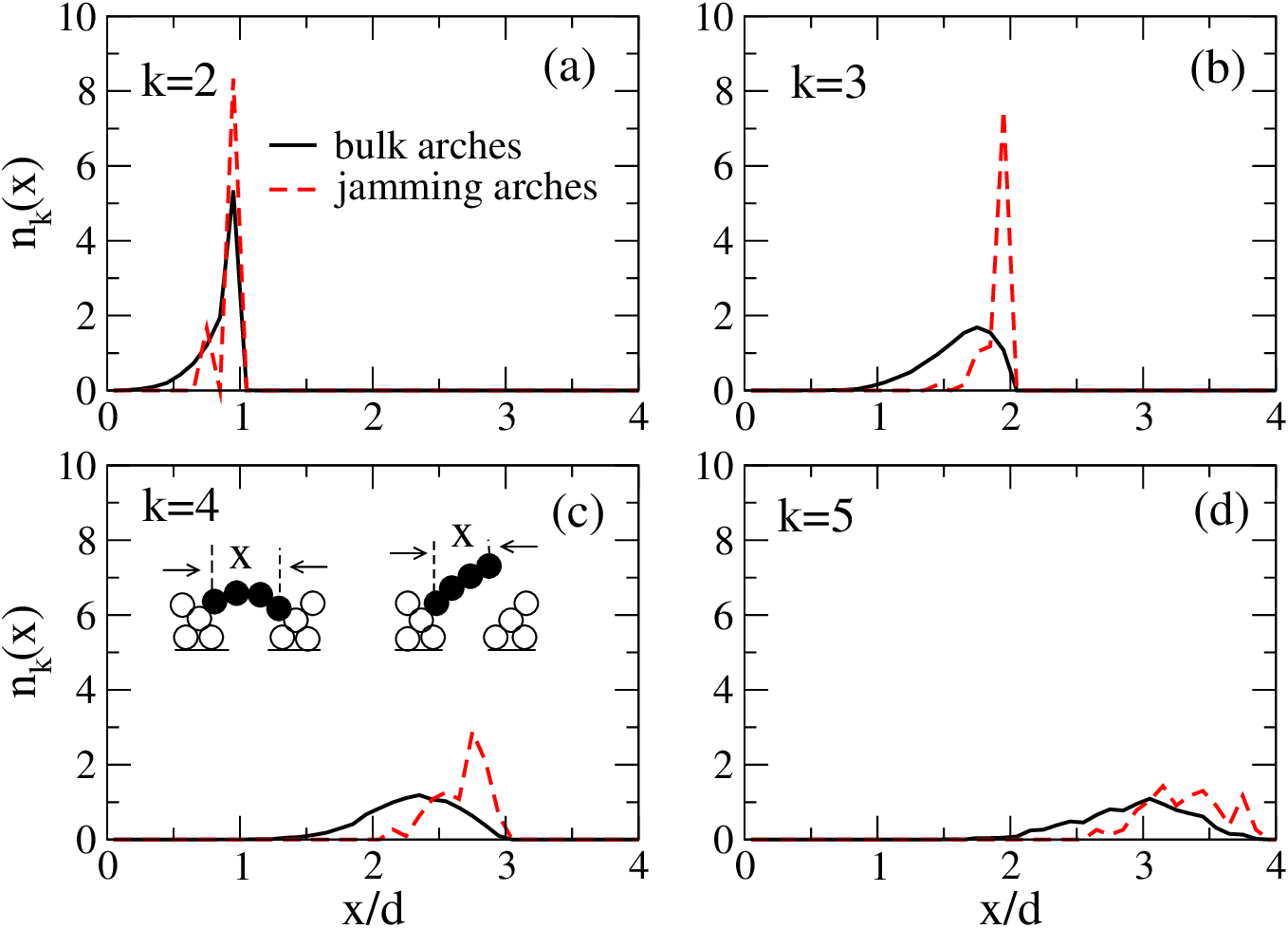}
\caption {(Color online). Horizontal span distribution $n_k(x)$ for arches formed by different number of grains $k$ at $\Gamma=0.707$. (a) $k=2$, (b) $k=3$, (c) $k=4$, (d) $k=5$. The red dashed lines correspond to blocking arches for an aperture $D=2.5$, whereas the black solid lines correspond to arches found in the bulk of the initial deposits. The inset of panel (c) shows two schematic arches of $k=4$ which are wide enough to jam the orifice but only one of them fits in the gap left by the two piles at rest on each side of the aperture.}
\label{fig8}
\end{center}
\end{figure}

The horizontal span of the arches for a given number of grains $k$ is shown in Fig.~\ref{fig8}. We include data for the bulk arches found in the initial deposits and for the jamming arches found for an aperture of width $D=2.5$. As we can see, for a given $k$, arches are more likely to be wider in the case of jamming arches as compared to bulk arches. This is to be expected for small $k$ since small arches with small span might not be able to jam the aperture. However, even arches with bigger $k$ are biased in the distribution of blocking arches. This is due to the two piles formed at each side of the orifice. Blocking arches that are wider than the aperture $D$ must span this funnel. Arches of $k$ disks with horizontal span sufficient to jam the orifice might still be unable to span the funnel (see the inset in Fig. \ref{fig8}(c)). This results in blocking arches generally wider, for a given $k$, than the corresponding bulk arches. Notice that arches of horizontal span $x<D-1$ may also jam the orifice (see Fig. \ref{fig8}(a)) due to the grains sitting at the edges of the aperture that reduce the effective size of the opening. This effect has been studied in more detail by Pournin et al. \cite{pournin}.   

\section{Conclusions}

We have considered granular avalanches discharged though small apertures at the bottom of a container by using a 2D pseudo-dynamic model. We have focused on the effect of the packing fraction of the granular deposit prior to the avalanche discharge. The results indicate that the initial packing fraction has an important effect on the mean avalanche size for a given opening size. However, similar $\langle s \rangle$ can be obtained for packings with very different $\phi$. Most importantly, very different values of $\langle s \rangle$ can correspond to initial packings with the same packing fractions that were prepared by using different tap intensities. It is important to note that these results are obtained not for single packings but for ensembles of deposits representative of steady states corresponding to a particular tap intensity. 

Our main conclusion is that packing fraction is not a good macroscopic parameter to predict the size of the avalanches that would flow through a given aperture. It seems that further information is necessary. Although this information is expected to reside in the size and number of arches, we have seen that  the correlation of these with $\langle s \rangle$ is not consistent for all openings $D$. It seems that is not possible to predict the jamming probability of a granular column as it flows through a small aperture based on a few global properties of the initial deposits. 

A side result from our study is that blocking arches are generally wider than the arches found in bulk. This is not only due to the fact that the aperture imposes a lower cut-off for the possible jamming arches, but also to the fact that even arches formed in the bulk which are wide enough to block the outlet have shapes not compatible with the conical boundary effectively created by the two piles of stationary grains at the sides of the aperture.

\ack
LAP acknowledges valuable discussion with Angel Garcimart\'{\i}n. This work has been partially supported by CONICET (Argentina).

\end{document}